\documentclass[aps,prd,10pt,nofootinbib,OliveGeqsecnum,showpacs,showkeys,superscriptaddress,preprintnumbers,altaffilletter,floatfix,twocolumn]{revtex4-2}

\usepackage[hidelinks]{hyperref}
\usepackage{graphicx,caption,subcaption,bm,color}

\usepackage{amsmath,amssymb}
\usepackage{mathtools} 
\usepackage{dsfont} 
\usepackage{multirow} 
\usepackage{float} 
\usepackage{mathrsfs} 
\usepackage{tensor} 
\usepackage{booktabs}
\usepackage{caption}
\usepackage{ragged2e}
\usepackage{mathrsfs}
\usepackage{upgreek}

\hypersetup{colorlinks, linkcolor={blue}, citecolor={red}, urlcolor={teal}}

\definecolor{amaranth}{rgb}{0.9, 0.17, 0.31}
\definecolor{palatinateblue}{rgb}{0.15, 0.23, 0.89}
\definecolor{brightpink}{rgb}{1.0, 0.0, 0.5}
\definecolor{brightgreen}{rgb}{0.14, 0.84, 0.72}
\definecolor{mediumgreen}{rgb}{0.22, 0.67, 0.59}
\definecolor{darkgreen}{rgb}{0.25, 0.5, 0.46}
\definecolor{verydarkgreen}{rgb}{0.22, 0.33, 0.31}

\hypersetup{ linktoc=all,
    colorlinks, linkcolor={mediumgreen},
    citecolor={mediumgreen}, urlcolor={darkgreen}
}
\DeclareUnicodeCharacter{03A9}{$\Omega$}
\DeclareUnicodeCharacter{039B}{$\Lambda$}

\graphicspath{{images/}}

\newcommand{\be}{\begin{equation}}
\newcommand{\ee}{\end{equation}}
\newcommand{\ba}{\begin{eqnarray}}
\newcommand{\ea}{\end{eqnarray}}

\begin{document}

\title{Cosmological perturbations for smooth sign-switching
dark energy models}

\author{Mariam Bouhmadi-López}
\email{mariam.bouhmadi@ehu.eus}
\affiliation{IKERBASQUE, Basque Foundation for Science, 48011, Bilbao, Spain}
\affiliation{Department of Physics  \& EHU Quantum Center, University of the Basque Country UPV/EHU, P.O. Box 644, 48080 Bilbao, Spain}

\author{Beñat Ibarra-Uriondo}
\email{benat.ibarra@ehu.eus}
\affiliation{Department of Physics \& EHU Quantum Center, University of the Basque Country UPV/EHU, P.O. Box 644, 48080 Bilbao, Spain}

\begin{abstract} 

In this work, we carry out a comprehensive perturbative analysis of four cosmological models featuring a sign-switching cosmological constant. Among these, we include the well-known $\Lambda_{\rm s}$CDM model, characterised by an abrupt transition from a negative to a positive cosmological constant. We also consider the L$\Lambda$CDM model, which exhibits a generalised ladder-step evolution, as well as the SSCDM and ECDM models, both of which undergo a smooth sign change at comparable redshifts. We solve the linear cosmological perturbation equations from the radiation-dominated era, imposing initial adiabatic conditions for matter and radiation, for modes well outside the Hubble radius in the early Universe. We analyse the behaviour of the matter density contrast, the gravitational potential, the linear growth rate, the matter power spectrum, and the $f\sigma_8$ evolution . These results are contrasted with predictions from the standard $\Lambda$CDM model and are confronted with observational data.

\end{abstract}


\maketitle


\section{Introduction}\label{intro}

First detected in 1998 from observations of SNeIa \cite{SupernovaSearchTeam:1998fmf}, the late-time expansion of the Universe has since been confirmed by various independent methods, including BAO, CMB, gravitational lensing, and many others \cite{Perlmutter_1999,Planck:2015bue,Demianski:2016dsa,Brout:2022vxf,DESI:2025zgx}.
The true nature driving this renewed inflationary phase remains unknown. Nevertheless, the cosmological constant continues to offer the simplest explanation for this accelerated expansion, giving rise to the concordance cosmology, namely the $\Lambda$CDM model. Although it remains the prevailing model, the standard cosmological framework faces several unresolved issues; most notably the $H_0$ \cite{DiValentino:2021izs,Poulin:2018,Kamionkowski:2022pkx} and $S_8$ tensions \cite{DiValentino:2021izs,Perivolaropoulos:2021jda,Abdalla:2022yfr}.

Numerous theoretical frameworks have been proposed to reconcile the observed discrepancies with $\Lambda$CDM, while retaining its remarkable success in explaining both early- and late-time cosmological observables \cite{Bamba:2012cp,CosmoVerse:2025txj}. Broadly, these proposals can be categorised into two main approaches. The first approach focuses on introducing alternative dark energy (DE) scenarios. These include early-time modifications to the Universe’s energy content, such as Early Dark Energy (EDE)\cite{Poulin:2018cxd}, New Early Dark Energy (NEDE)\cite{Niedermann:2019olb,Cruz:2023lmn,Niedermann:2023ssr}, and Anti-de Sitter Early Dark Energy (AdS-EDE)\cite{Ye:2021iwa,Ye:2020btb,Ye:2020oix}. Late-time alternatives, which adjust the matter content of the Universe through various DE formulations, have also been proposed. Examples include phantom-crossing DE models\cite{DiValentino:2020naf,Alestas:2020mvb}, the Omnipotent DE scenario~\cite{DiValentino:2020naf,Adil:2023exv}, (non-minimally) interacting DE models~\cite{Kumar:2017dnp,Nunes:2022bhn}, and  axion-like DE models~\cite{Kamionkowski:2014zda,Emami:2016mrt,Chiang:2025qxg}. The second  involves modifications to the theory of gravity itself, aiming to account for the late-time accelerated expansion of the Universe without invoking a conventional DE component. These Modified Gravity models extend General Relativity (GR) by altering the Einstein–Hilbert action—either through the inclusion of additional scalar degrees of freedom, modifications to the underlying geometric structure, or changes in spacetime dimensionality, among other possibilities \cite{CANTATA:2021asi}. Such modifications can affect both the background expansion history and the evolution of large-scale structure. Notable examples include $f(\mathcal{R})$ gravity~\cite{Sotiriou:2008rp,DeFelice:2010aj,Nojiri:2010wj,Nojiri:2017ncd}, $f(\mathcal{T})$ gravity~\cite{Bengochea:2008gz,Ferraro:2006jd,Cai:2015emx}, $f(\mathcal{R},T)$ models~\cite{Harko:2011kv,Nojiri:2004bi}, $f(\mathcal{Q})$ theories~\cite{BeltranJimenez:2018vdo,BeltranJimenez:2019tme,Ayuso:2020dcu,Boiza:2025xpn,Ayuso:2025vkc}, bimetric gravity~\cite{Kobayashi:2019hrl}, and kinetic gravity braiding (KGB)~\cite{Deffayet:2010qz,Pujolas:2011he,BorislavovVasilev:2022gpp,BorislavovVasilev:2024loq} among many others.

Within the first framework we find the $\Lambda_{\rm s}$CDM model, originally proposed in \cite{Akarsu:2021fol} and based on the findings of the graduated dark energy (gDE) model \cite{Akarsu:2019hmw,Acquaviva:2021jov}, which extends the concordance cosmology by replacing the positive cosmological constant (CC) with a dynamically evolving counterpart that undergoes a sign-switching transition from negative to positive values at low redshift.  This transition has thus far been modelled using sigmoid-like functions, with the extreme case being the signum function. Smoother variants have also been explored, such as the hyperbolic tangent function \cite{Akarsu:2025gwi}. Although rapid AdS-to-dS-like transitions in DE were initially considered difficult to justify through a well-founded physical mechanism, the remarkable phenomenological success of the $\Lambda_{\rm s}$CDM model \cite{Akarsu:2025dmj, Escamilla:2025imi, Souza:2024qwd, Yadav:2024duq, Akarsu:2024eoo, Akarsu:2023mfb, Akarsu:2022typ,Akarsu:2021fol}, despite its simplicity, has rekindled theoretical interest. Recent research has shown that even well-established theories, when revisited, can reveal previously unexplored solution spaces that naturally accommodate such transitions. This compels a re-evaluation of traditional theoretical paradigms.
A notable example is the $\Lambda_{\rm s}$CDM$^+$ model \cite{Anchordoqui:2023woo,Anchordoqui:2024gfa,Anchordoqui:2024dqc,Soriano:2025gxd} , which builds upon the $\Lambda_{\rm s}$CDM framework within the context of string theory. In the $\Lambda_{\rm s}$CDM$^+$ model the negative to positive DE density transition is realised through the action of Casimir forces in the bulk; i.e. the higher dimensional space-time. Furthermore, frameworks involving Barrow entropy \cite{DiGennaro:2022ykp}, two interacting fluid models \cite{Ong:2022wrs}, quintessence models with early DE-like features \cite{Andriot:2025los}, the possibility that $\Omega_{\textrm{m}}>1$, inducing negative DE densities \cite{Malekjani:2023ple} and even bimetric gravity models \cite{Dwivedi:2024okk} have also been employed to describe sign-switching models.

Even though the abrupt $\Lambda_{\rm s}$CDM scenario, regarded as the simplest phenomenological realisation of the $\Lambda_{\rm s}$CDM framework, has been extensively studied \cite{Akarsu:2019hmw,Akarsu:2021fol,Akarsu:2022typ,Akarsu:2023mfb} under the assumption that its dynamics is governed by GR, the model’s instantaneous transition introduces a discontinuity at $z = z_\dagger$, resulting in a type II (sudden) singularity \cite{Nojiri:2005sx}. Despite this, a detailed analysis in \cite{Paraskevas:2024ytz} has demonstrated that the impact of this singularity on the formation and evolution of cosmic structures is negligible. Furthermore, it was shown in \cite{Bouhmadi-Lopez:2025ggl} that continuous DE density transitions give rise to $w$-singularities, which are milder in nature than sudden singularities \cite{Bouhmadi-Lopez:2019zvz}. Nonetheless, the presence of a sudden singularity implies that the abrupt $\Lambda_{\rm s}$CDM model should be interpreted as a phenomenological approximation; i.e. an idealised representation of a transition which, in reality, is likely to be rapid but smooth. A physically well-motivated formulation necessitates a continuous transition, permitting a comprehensive analysis of its dynamics and its influence on cosmic perturbations. However, the incorporation of a smooth transition presents additional theoretical challenges that must be addressed with care. 

In Ref.~\cite{Bouhmadi-Lopez:2025ggl}, we introduced three novel extensions of the $\Lambda_{\rm s}$CDM model: one featuring a ladder-like cosmological constant (L$\Lambda$CDM), another inspired by a smooth step function (SSCDM), and a third following the profile of an error function (ECDM). These models, whose background evolution have already been analysed, appear to be well suited extensions of the abrupt cosmological constant sign-switching scenario, exhibiting smoother behaviour. However, to fully assess their viability, one must move beyond background-level analysis and investigate their implications for structure formation. For this purpose, perturbation theory offers a robust mathematical framework to examine cosmic inhomogeneities and extract observable quantities, such as the distribution of $f\sigma_8$, enabling direct comparison between theoretical predictions and observational data on matter clustering. Although perturbations have been extensively studied in the literature \cite{Mukhanov:1990me,Wang:2009azb,Malik_2009,Boiza:2024fmr}, it suffices here to adopt a multi-fluid approach, as employed in \cite{Albarran:2016mdu}.

This paper is organised as follows. In Sec.~\ref{sec2}, we present and review the DE models we will consider on this work. In Sec.~\ref{sec3}, we introduce the general framework for studying scalar linear perturbations. In Subsec.~\ref{sec3a}, we discuss the initial conditions chosen to numerically solve the perturbation equations in our
model. Our numerical results are presented in Sec.~\ref{sec4}, where we examine the evolution of matter perturbations, $\delta_m$, and the gravitational potential, $\Psi$. In Sec.~\ref{observables}, we investigate structure formation using several measurements of the growth rate function, specifically, $f\sigma_8$, alongside the matter power spectrum. Finally, in Sec.~\ref{sec5}, we conclude by highlighting the key aspects of our study.

\section{Sign-switching DE\label{sec2}}

In this section, we briefly review  the DE models considered in  the present work, which 
we shall refer to as: (A) abrupt sign-switching DE ($\Lambda_{\rm s}$CDM), (B) Ladder-like DE (L$\Lambda$CDM), (C) Smooth step DE (SSCDM) (D) Error DE (ECDM). 
For each model, we begin by presenting DE density evolution and the equation of state (EoS) for DE that gives rise to the model, ensuring that the background evolution follows closely that of $\Lambda$CDM until the present time. 

We consider a cosmological background with spatial geometry described by the Friedmann–Lemaître–Robertson–Walker (FLRW) metric:
\begin{equation}
d s^2 = - dt^2 + a^2 \delta_{ij} \, dx^i dx^j,
\label{gauge0}
\end{equation}
where $ a(t) $ is the scale factor, $ t $ denotes cosmic time, and $ \delta_{ij} $ is the spatial metric. We examine a Universe comprising three principal components: radiation, non-relativistic matter (baryons and dark matter), and DE. Each fluid is modelled as a perfect fluid with energy density $ \rho_A $ and pressure $ p_A = w_A \rho_A $, where the subscript $ A $ corresponds to radiation ($ r $) with $ w_\mathrm{r} = 1/3 $, non-relativistic matter ($ m $) with $ w_\mathrm{m} = 0 $, and DE ($ d $) with a redshift-dependent EoS $ w_\mathrm{d} = w_\mathrm{d}(z) $.

Assuming conservation of the energy–momentum tensor for each component, we obtain:
\begin{equation}
    \dot{\rho}_A + 3H\left(\rho_A + p_A\right) = 0.
\end{equation}

It follows that the dark energy density evolves with redshift as:
\begin{equation}
    \rho_{\mathrm{d}}(z) = \rho_{\mathrm{d},0} \exp\left[3 \int_0^z \frac{1 + w_\mathrm{d}(z')}{1 + z'} \, dz'\right].
\end{equation}

The Friedmann equation for a spatially flat universe reads:
\begin{equation}
    \frac{H^2}{H_0^2} = \Omega_{\mathrm{r},0}(1 + z)^4 + \Omega_{\mathrm{m},0}(1 + z)^3 + \Omega_{\mathrm{d},0} \frac{\rho_{\mathrm{d}}(z)}{\rho_{\mathrm{d},0}},
\end{equation}
where $ \Omega_{A,0} \equiv \frac{8\pi G }{3H_0^2}\rho_{A,0} $ represents the present-day fractional energy density of component $A$, and $ H \equiv \dot{a}/a $ is the Hubble parameter. The notation $ \dot{\{ \text{ }\}} \equiv d/dt $ denotes differentiation with respect to cosmic time. Throughout this work, a subscript ‘0’ denotes present-day values. The fractional energy densities satisfy the closure condition:
$
1 = \Omega_{\mathrm{d},0} + \Omega_{\mathrm{m},0} + \Omega_{\mathrm{r},0}.
$

The total EoS parameter can also be expressed in terms of the fractional energy densities as:
\begin{equation}
    w \equiv \frac{p}{\rho} = \Omega_\mathrm{r} w_\mathrm{r} + \Omega_\mathrm{m} w_\mathrm{m} + \Omega_\mathrm{d} w_\mathrm{d}.
\end{equation}

In the subsections that follow, we first review the example model previously studied in~\cite{Akarsu:2021fol}, and subsequently introduce the three new models developed in~\cite{Bouhmadi-Lopez:2025ggl}.

\subsection{Abrupt sign-switching DE (\texorpdfstring{$\Lambda_{\rm s}$}{Lambdas}CDM)}
This model is a one parameter  extension of $\Lambda$CDM  where the extra parameter is $z_{\dagger,s}$, the redshift at which the cosmological constant changes sign
    \begin{equation}
        \Lambda_\mathrm{s}(z)=\Lambda  \text{ sgn}(z_{\dagger,s}-z).
    \end{equation}
    As DE density remains constant over time, the EoS parameter remains constant at all time, except for the sign-switching:
\begin{equation}
    w_{\textrm{d,s}}=-1-\frac{\updelta(z_{\dagger,s}-z)}{3 \textrm{ sgn}(z_{\dagger,s}-z)}(1+z).
\end{equation}

Here, $\updelta(z_{\dagger,s} - z)$ stands for the Dirac delta function. It can be shown that a sudden singularity arises at the sign-switching point \cite{Barrow1,Bouhmadi-Lopez:2019zvz,Paraskevas:2024ytz}. It was shown in \cite{Paraskevas:2024ytz} that these singularities are relatively mild. Indeed, no bounded structures are destroyed. For further details on this model, please refer to Ref. \cite{Akarsu:2021fol}.

\subsection{Ladder-like DE (L\texorpdfstring{$\Lambda$}{Lambda}CDM)}

In this model, the effective cosmological constant increases gradually in discrete steps. Beginning with a negative DE density in the past, the density increases through a series of small discontinuous jumps, which we have set as $N=20$. The height of each rung depends on the total number of steps assumed. For simplicity, we assume that all rungs span equal intervals in redshift.
 Thus, this model represents a two-parameter extension of the standard $\Lambda$CDM framework. The additional parameters are: the sign-switching redshift, $z_{\dagger,l}$; and the step length, $\Delta z_\mathrm{step}$. These parameters collectively determine the initial and final redshifts at which the ladder-step transition occurs, given by
$z_{i,l} = z_{\dagger,l} + \Delta z_\mathrm{step}N/2$ and $z_{f,l} = z_{\dagger,l} - \Delta z_\mathrm{step}N/2$, respectively\footnote{There are two possible configurations for ladder-like DE models, depending on whether the number of steps $N$ is even or odd, as discussed in Ref.~\cite{Bouhmadi-Lopez:2025ggl}. This choice affects the precise values of the initial and final redshifts of the transition. However, we focus here exclusively on the even case, as both configurations yield indistinguishable results at the perturbative level.}. The DE density takes the form:

\begin{equation}
      \begin{aligned}
     &  \Lambda_\mathrm{l}(z)=\Lambda \left[1-\frac{2}{N}\sum_{n=1}^N \mathscr{H}
(z_n-z)\right],
     \end{aligned}  
\end{equation}
where $\mathscr{H}
(z_n - z)$ is the Heaviside step function evaluated at $z_n - z$ and $z_n=z_{f,le}+\frac{n}{N}(z_{i,le}-z_{f,le})$ denotes the redshift at which the transition between steps occurs with $n \in [0, N]$ representing the step index at that time, ordered from present to past. The total EoS parameter, $w$, also exhibits discontinuities with each step transition, giving rise to sudden singularities
\cite{Bouhmadi-Lopez:2025ggl}:

\begin{equation}
    w_{\textrm{d,l}}=-1-\frac{(1+z)\sum^{N}_{i=1}\updelta(z_n-z)}{3 \left[-\frac{N}{2}+\sum^{N}_{n=1}\mathscr{H}
(z_n-z)\right]},
\end{equation}
where $\updelta(z_n - z)$ stands for the Dirac delta function.
\subsection{Smooth step DE (SSCDM)}
This model constitutes a two-parameter extension of the $\Lambda$CDM framework, characterised by a smooth transition of DE density from negative to positive values . The additional parameters, $z_{i,ss}$ and $z_{f,ss}$, correspond to the redshifts at which this transition begins and ends, respectively:

\begin{equation}
  \begin{aligned}
&  \Lambda_\mathrm{ss}(x)=\Lambda \begin{cases}-1, & x \le x_{i,ss},\\  1-2(126 t^5 - 420 t^6 \\
+ 540 t^7 - 315 t^8 + 70 t^9),  & x_{i,ss}<x<x_{f,ss},\\
1, & x\ge x_{f,ss},\end{cases} \\
\end{aligned}  
\label{dnesitysscdm}
\end{equation}
where $t=\frac{x-x_{f,ss}}{x_{i,ss}-x_{f,ss}}$ and $x=\ln(a/a_0)$. In this case the EoS parameter will also be equal to -1 in most of the regime, except during the transition period,  between $x_{i,ss}$ and $x_{f,ss}$, during which

\begin{equation}
  \begin{aligned}
&  w_\mathrm{d,ss}(x)= \begin{cases}-1, & x \le x_{i,ss},\\-1  -\frac{1260}{\Delta x \text{ }\Lambda_\mathrm{ss}(x)/\Lambda}(t^4 \\-4t^5+6t^6-4t^7+t^8),  & x_{i,ss}<x<x_{f,ss},\\
,-1 & x\ge x_{f,ss},\end{cases}\\
\end{aligned}  
\label{eossscdm}
\end{equation}

\subsection{Error-like DE (ECDM)}

This model represents another two-parameter extension of the $\Lambda$CDM framework, in which the DE density transitions from negative values at high redshift to positive values at low redshift via an error function. The transition is characterised by $\eta$, which determines the steepness of the curve during the transition epoch, and $z_{\dagger,e}$, the redshift at which the DE density changes sign. The DE energy density can be written as:

\begin{equation}
       \Lambda_\mathrm{e}(x)=\Lambda_{ } \text{Erf}\left[\eta(x-x_{\dagger,e})\right].
\end{equation}
The parameter $x_{\dagger,e}$ is defined as  $x_{\dagger,e} = -\ln(1 + z_{\dagger,e})$.

The EoS parameter evolves over time in accordance with the conservation of the DE density

\begin{equation}
    w_\mathrm{d,e}(x)=-1-\frac{2\eta e^{-\eta^2(x-x_{\dagger,e})^2}}{3\sqrt{\pi}\text{Erf}[\eta(x-x_{\dagger,e})]}.
    \label{eosecdm}
\end{equation}

In the case of the smooth-transitioning models, (C) and (D), 
$w$-singularities emerge at the sign-switching point \cite{Bouhmadi-Lopez:2025ggl}. However, these singularities are milder than the so-called sudden singularities \cite{Fernandez-Jambrina:2010ngm,Fernandez-Jambrina:2006tkb,Bouhmadi-Lopez:2019zvz} . For further details on models (B), (C) and (D) please refer to Ref. \cite{Bouhmadi-Lopez:2025ggl}.

\section{COSMOLOGICAL PERTURBATIONS \label{sec3}}

Cosmological perturbations have been extensively analysed in the literature \cite{Brandenberger:1993zc,Ma:1995ey,2004,Malik_2009,bauman}. In this section, we introduce the Newtonian gauge and employ the corresponding gauge-invariant quantities to compute perturbations up to first order. The perturbed FLRW metric in the Newtonian gauge is given by \cite{bauman}:

\begin{equation}
d s^2=a^2\left[-\left(1+2 \Phi\right) d \tau^2+\left(1-2 \Psi\right)\delta_{ij} d x^i dx^j\right],
\label{gauge1}
\end{equation}
where $\tau$ is the conformal time, defined by $d\tau = dt / a$, and $\Phi\left(\eta, x^i\right)$ and $\Psi\left(\eta, x^i\right)$ are the scalar metric perturbations that correspond to the Bardeen potentials \cite{PhysRevD.22.1882}. 
It can be shown that the perturbed Einstein equations, encapsulated in a system of three equations governing the gravitational potential and its derivative, are given by \footnote{In the absence of the anisotropic stress tensor, the gravitational potentials are identical, such that $\Phi=\Psi$. A detailed proof of this statement can be found, for example, on \cite{Albarran:2016mdu}.}:

\begin{equation}
    \begin{aligned} 
    3 \mathcal{H}\left(\mathcal{H} \Psi+\Psi^{\prime}\right)-\nabla^2 \Psi & =-\frac{\kappa^2a^2}{2} \delta \rho, \\ 
    \nabla^2\left(\mathcal{H} \Psi+\Psi^{\prime}\right) & =-\frac{\kappa^2a^2}{2}(\rho+p) \nabla^2 v, \\ 
    \Psi^{\prime \prime}+3 \mathcal{H} \Psi^{\prime}+\Psi\left(2 \mathcal{H}^{\prime}+\mathcal{H}^2\right) & =\frac{\kappa^2a^2}{2} \delta p.
    \end{aligned}
    \label{field}
\end{equation}

We will follow the multi-fluid approach used in \cite{Boiza:2024fmr,Albarran:2020bwn,PhysRevD.22.1882}. The total energy density perturbation, $\delta\rho$, the total pressure perturbation, $\delta p$, and the total peculiar velocity perturbation, $\delta v$,  can be expressed in terms of the contributions from each individual fluid \footnote{$\delta$ denotes a perturbative variable, not to be confused with the Dirac delta function $\updelta(z)$.}: 
\begin{equation}
    \delta \rho=\sum_A\delta \rho_A, \quad \delta p=\sum_A\delta p_A, \quad v=\sum_A\frac{1+w_A}{1+w}\Omega_A v_A.
\end{equation}

The perturbation of the conservation of the energy-momentum tensor, $T^\mu_{A \text{ }\nu}$, for each fluid $A$ yields the evolution equations for the fractional energy density perturbation, $\delta_A = \delta \rho_A / \rho_A$, and the peculiar velocity potential, $v_A$:
\begin{equation}
    \begin{aligned}
        \delta_A^{\prime}=  3 \mathcal{H}\left(w_A-c_{s A}^2\right) \delta_A+3\left(1+w_A\right) \Psi^{\prime} \\
        +\left(1+w_A\right)\left[9 \mathcal{H}\left(c_{s A}^2-c_{a A}^2\right)-{\nabla^2}\right] v_A ,
    \end{aligned}
    \label{dengen}
\end{equation}
\begin{equation}
    v_A^{\prime}= \left(3 c_{s A}^2-1\right) \mathcal{H} v_A-\frac{c_{s A}^2}{\left(1+w_A\right)} \delta_A-{\Psi},
    \label{velgen}
\end{equation}
where a prime stands for a derivative with respect to the conformal time and $\mathcal{H}$ is the conformal Hubble parameter, $\mathcal{H}=$ $a H$. We have introduced the adiabatic speed, $c_{a A}^2$,  and the rest frame sound speed, $c_{s A}^2$, for each fluid A. These are respectively given by 
\begin{equation}
    c_{a A}^2=\frac{p_A^{\prime}}{\rho_A^{\prime}}, \quad
    c_{s A}^2=\left.\frac{\delta p_A}{\delta \rho_A}\right|_{\mathrm{r.f.}}.
\end{equation}

In order to study the evolution of linear perturbations, we move to Fourier space, where we decompose the functions $\psi(\tau,x^i)$ into its Fourier components $\psi_k(\tau)$:

\begin{equation}
    \psi(\tau,\boldsymbol{x})=\frac{1}{(2\pi)^{3/2}}\int e^{-i\boldsymbol{k}\cdot \boldsymbol{x}}\psi_k(\tau) d^3\boldsymbol{k}.
\end{equation}

By applying  a Fourier decomposition to  Eqs. (\ref{field}), (\ref{dengen}) and (\ref{velgen}), we obtain the evolution equations for the fractional energy density perturbation and the peculiar velocity potential, for each mode and for the different components of the universe (radiation, dust and DE): 
\begin{equation}
\begin{aligned}
 \left(\delta_{\mathrm{r}}\right)_x=&\frac{4}{3}\left(\frac{k^2}{\mathcal{H}} v_{\mathrm{r}}+3 \Psi_x\right), \\
 \left(v_r\right)_x=&-\frac{1}{\mathcal{H}}\left(\frac{1}{4} \delta_{\mathrm{r}}+\Psi\right), \\
\left(\delta_{\mathrm{m}}\right)_x=&\left(\frac{k^2}{\mathcal{H}} v_{\mathrm{r}}+3 \Psi_x\right), \\
 \left(v_{\mathrm{m}}\right)_x=&-\left(v_{\mathrm{m}}+\frac{\Psi}{\mathcal{H}}\right), \\
\left(\delta_{\mathrm{d}}\right)_x=&\left(1+w_{\mathrm{d}}\right)\left\{\left[\frac{k^2}{\mathcal{H}}+9 \mathcal{H}\left(c_{ \mathrm{sd}}^2-c_{\mathrm{ad}}^2\right)\right] v_{\mathrm{d}} \right.\\
& \left. +3 \Psi_x\right\}  +3\left(w_{\mathrm{d}}-c_{ \mathrm{sd}}^2\right) \delta_{\mathrm{d}}, \\
 \left(v_{\mathrm{d}}\right)_x=&-\frac{1}{\mathcal{H}}\left(\frac{c_{ \mathrm{sd}}^2}{1+w_{\mathrm{d}}} \delta_{\mathrm{d}}+\Psi\right)+\left(3 c_{ \mathrm{sd}}^2-1\right) v_{\mathrm{d}},
\end{aligned}
\label{perturb}
\end{equation}

and for the metric potential

\begin{equation}
\begin{aligned}
&\Psi_x+\Psi\left(1+\frac{k^2}{3 \mathcal{H}^2}\right)  =-\frac{1}{2} \delta, \\
&\Psi_x+\Psi  =-\frac{3}{2} \mathcal{H} v(1+w), \\
&\Psi_{x x}+\left[3-\frac{1}{2}(1+3 w)\right] \Psi_x-3 w \Psi  =\frac{3}{2} \frac{\delta p}{\rho},
\label{fieldk}
\end{aligned}
\end{equation}
where $\{\,\}_x$ denotes derivatives with respect to $\ln(a/a_0)$. 

Regarding DE perturbations, these do not arise in models (A) and (B), since the DE density remains constant over time. For models (C) and (D), however, while the DE density evolves, it is reasonable to neglect the associated perturbations in sufficiently rapid transitions, as the variation in DE is confined to a very short timescale.

\subsection{Initial conditions \label{sec3a}}
We need to set proper initial conditions to compute the evolution of the perturbations (cf.~\cite{Ma:1995ey,Ballesteros:2010ks} for a detailed discussion regarding initial conditions on DE models). We impose our initial conditions during the radiation-dominated era, around $z\sim 10^6$. At this moment, all the relevant modes in the linear regime are super-Hubble, with wave numbers satisfying
 $k\ll \mathcal{H}$. This means they lie outside the horizon. Taking into account this two approximations, we combine the first and third equations in Eq. (\ref{fieldk}) obtaining
\begin{equation}
    \Psi_{xx}+3\Psi_x\approx 0.
    \label{approx}
\end{equation}
The dominant solution to Eq. (\ref{approx})
is a constant solution $\Psi_{ini}=\Psi(x_{ini})$. Applying this result to the set of Eqs.(\ref{fieldk}), we obtain

 \begin{equation}
     \begin{aligned} & \Psi_{\textrm{ini}}=-\frac{\delta_{\textrm{ini}}}{2\left[1+k^2 /\left(3 \mathcal{H}_{\textrm{ini}}^2\right)\right]}, \\ & \Psi_{\textrm{ini}}=-\frac{3}{2} \mathcal{H}_{\textrm{ini}} v_{\textrm{ini}}\left(1+w_{\textrm{ini}}\right).\end{aligned}
     \label{approx2}
 \end{equation}

In addition, motivated by single scalar field models of inflation \cite{Ma:1995ey,Ballesteros:2010ks,Amendola:2015ksp}, we adopt adiabatic initial conditions, given by 
\begin{equation}
    \frac{\delta_\textrm{r}}{1+w_\textrm{r}}= \frac{\delta_\textrm{m}}{1+w_\textrm{m}}=  \frac{\delta}{1+w},
    \label{iniper}
\end{equation}
\begin{equation}
v_{\textrm{r,ini}}=v_{\textrm{m,ini}}=v_{\textrm{ini}}.
    \label{inivel}
\end{equation}

By combining Eqs. (\ref{iniper}) and (\ref{inivel}) with Eq.~(\ref{approx2}), we derive the initial conditions for both the perturbations and their associated velocities.
\begin{equation}
    \begin{aligned}
        \frac{3}{4} \delta_{\textrm{r,ini}}= \delta_{\textrm{m,ini}}\approx \frac{3}{4} \delta_{\textrm{ini}}  =
        -\frac{2[1+k^2/(3\mathcal{H}_{\textrm{ini}}^2)]\Psi_{\textrm{ini}}}{1+w_{ini}},
    \end{aligned}
    \label{ini1}
\end{equation}
\begin{equation}
    v_{\textrm{r,ini}}=v_{\textrm{m,ini}}=-\frac{2 \Psi_{\textrm{ini}}}{3 \mathcal{H}_{\textrm{ini}} \left(1+w_{\textrm{ini}}\right)}.
    \label{ini2}
\end{equation}


Taking advantage of the linearity of Eqs.~(\ref{dengen}) and (\ref{velgen}), we first compute the evolution of the perturbation quantities using the initial conditions (\ref{ini1}) and (\ref{ini2}) with $\Psi_{\text{ini}} = 1$ (which implies $\delta_{\text{ini}} = -2$), and subsequently rescale all resulting solutions by the physical value of $\delta_{\text{phys}}(k)$ \cite{Albarran:2016mdu}. The latter is taken from the Planck observational fit to single-field inflation \cite{Planck:2018vyg}:

\begin{equation}
\delta_{\text {phys }}(k)=\frac{8 \pi}{3} \sqrt{2 A_s}\left(\frac{k}{k_{\text {pivot }}}\right)^{\frac{n_s-1}{2}} k^{-\frac{3}{2}} .
\end{equation}
Here, $A_s$ and $n_s$ are defined as the amplitude and spectral index of the primordial inflationary power spectrum corresponding to a previously selected pivot scale $k_{\text {pivot }}$. In this work, we will use the values $k_{\text {pivot }}=0.05 \text{ h Mpc}^{-1}$, $A_s=2.101 \times 10^{-9}$, and $n_s=0.9665$ in accordance with Planck observational data \cite{Planck:2018vyg}.

\section{NUMERICAL RESULTS \label{sec4}}

\begin{figure*}[htbp] 
    \centering
    \begin{subfigure}{0.45\textwidth}
        \centering
        \includegraphics[width=\textwidth]{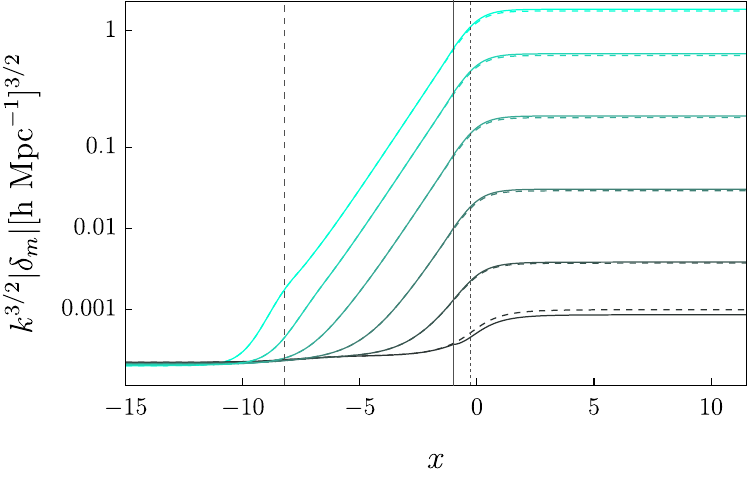}
        \caption*{(A) $\Lambda_{\rm s}$CDM}
    \end{subfigure}
    \hspace{0.05\textwidth}
    \begin{subfigure}{0.45\textwidth}
        \centering
        \includegraphics[width=\textwidth]{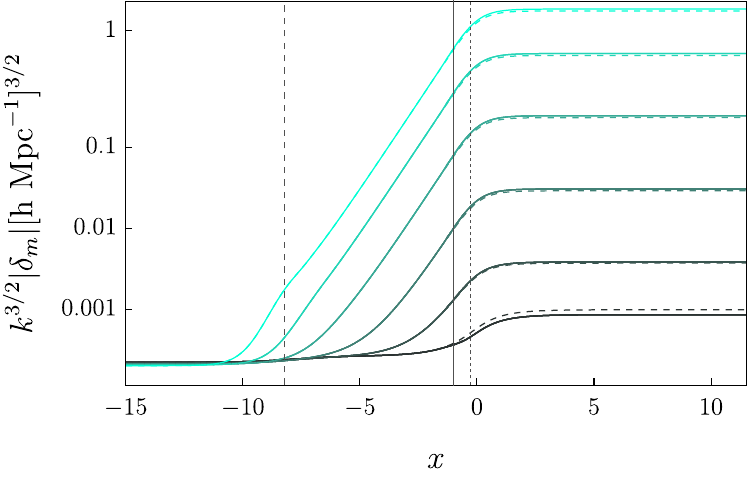}
        \caption*{(B) L$\Lambda$CDM}
    \end{subfigure}

    \vspace{0.5cm} 

    \begin{subfigure}{0.45\textwidth}
        \centering
        \includegraphics[width=\textwidth]{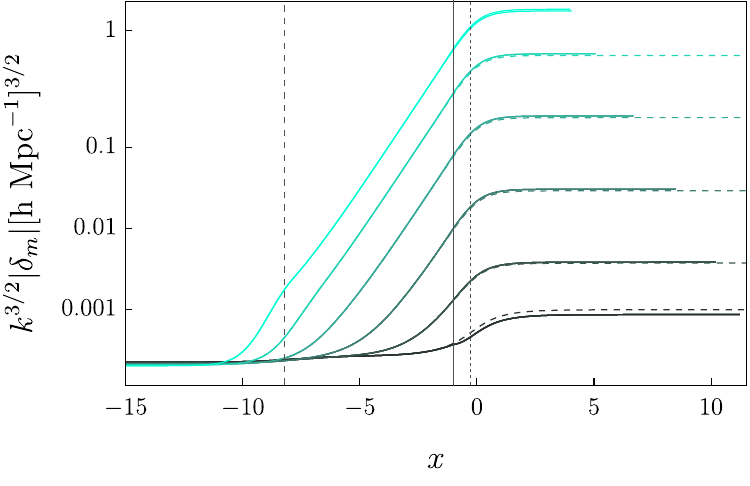}
        \caption*{(C) SSCDM}
    \end{subfigure}
    \hspace{0.05\textwidth}
    \begin{subfigure}{0.45\textwidth}
        \centering
        \includegraphics[width=\textwidth]{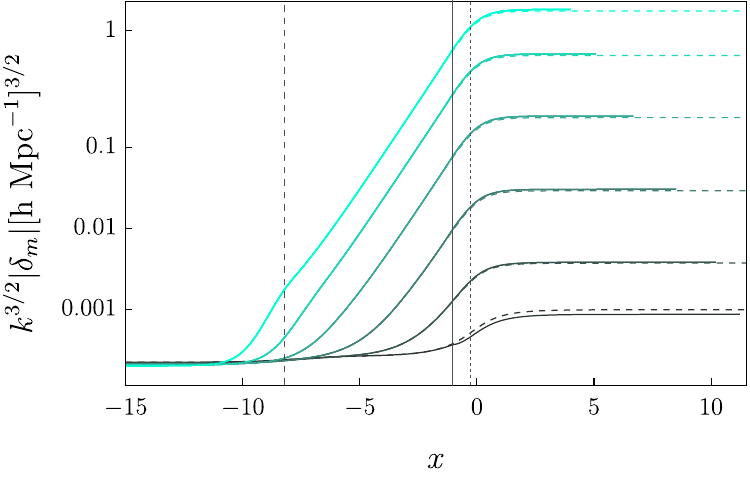}
        \caption*{(D) ECDM}
    \end{subfigure}
    \caption[Matter perturbations comparison]{\justifying{\textit{
Evolution of matter perturbations, $\delta_m$, for the four sign-switching models, compared with the standard $\Lambda$CDM scenario. In all four cases, the behaviour of the perturbations closely resembles that of $\Lambda$CDM, rendering the models largely indistinguishable in this respect. In each panel, solid lines correspond to the model under consideration, while dotted lines represent the $\Lambda$CDM predictions. Different colours denote distinct Fourier modes:  
$k = 3.33 \times 10^{-4} \ \text{h} \ \text{Mpc}^{-1}$ (Dark Gray-Green),  
$k = 1.04 \times 10^{-3} \ \text{h} \ \text{Mpc}^{-1}$ (Deep Green),  
$k = 3.27 \times 10^{-3} \ \text{h} \ \text{Mpc}^{-1}$ (Dark Teal),  
$k = 1.02 \times 10^{-2} \ \text{h} \ \text{Mpc}^{-1}$ (Turquoise),  
$k = 3.19 \times 10^{-2} \ \text{h} \ \text{Mpc}^{-1}$ (Teal Green),  
$k = 0.1 \ \text{h} \ \text{Mpc}^{-1}$ (Aqua Blue).  The perturbations are plotted as a function of $x = \ln(a/a_0)$, ranging from the radiation-dominated era ($x = -15$) to the far future ($x = 12$), with $x = 0$ corresponding to the present time. The left dashed vertical line marks the radiation–matter equality, the right dashed line indicates the matter–DE equality, and the solid vertical line denotes the redshift at which the DE density changes sign.
}} }

    \label{fig:four-images}
\end{figure*}
In this section, we present a discussion of the results obtained from the evolution of cosmological perturbations for the models introduced in Sec.~\ref{sec2}, along with their comparison to the $\Lambda$CDM model. For each case, we compute the evolution of the perturbation variables $\delta_m$, $v_m$, $\delta_r$, and $v_r$ by numerically integrating the system of Eqs.~(\ref{dengen}) and (\ref{velgen}), after substituting the gravitational potential $\Psi$ and its derivative $\Psi_x$ as given by Eq.~(\ref{fieldk}). The integration is performed from deep within the radiation-dominated epoch ($z \sim 10^6$), when all relevant modes lie outside the horizon, up to a point in the far future ($z \sim -0.99$). Initial conditions are set in accordance with Eqs.~(\ref{ini1}) and (\ref{ini2}). The integration is repeated for modes ranging from $k = 0.1\, \text{h}\, \text{Mpc}^{-1}$ down to $k = 3.33 \times 10^{-4}\, \text{h}\, \text{Mpc}^{-1}$. The cosmological parameters $\Omega_{m,0}$, $\Omega_{r,0}$, and $H_0$ used across all models are taken from the Planck mission results \cite{Planck:2018vyg}. Moreover, we must specify the parameters required to define each particular model. This is achieved as follows: 

\begin{itemize}
    \item Model (A): $z_{\dagger,s}$=1.7 according to \cite{Akarsu:2022typ,Akarsu:2023mfb,Paraskevas:2024ytz}. 
    \item Model (B): By selecting the parameter values $N=20$, $z_{\dagger,l}=2.7$ and  $\Delta z_{\mathrm{step}} = 0.1$, we set the initial and final redshifts of the transition to 
    \footnote{This is an approximated ``fit'' based on the values obtained in \cite{Akarsu:2025ijk,Escamilla:2025imi}, on the sense that we took the values of $H_0$, $z_\dagger$, $\eta$, $\Omega_{\mathrm{m},0}$, $\Omega_{\mathrm{r},0}$ and $\Lambda$ and adjusted them to our model. We are currently observationally fitting properly these models and we will report soon our results.}  to $z_{l,i} = 0.7$ and $z_{l,f} = 1.7$, respectively.
    \item Model (C): The values $z_{ss,f} = 0.6$ and $z_{ss,i} = 2.6$ were chosen to reproduce a level of smoothness comparable to that of Model (B), while ensuring that the sign-switch occurs at $z_{\dagger,ss} = 1.72$, allowing for a direct comparison with Model\footnote{It should be noted that reducing the separation between $z_{ss,i}$ and $z_{ss,f}$ renders the transition increasingly abrupt, effectively approximating a model similar to $\Lambda_{\rm s}$CDM.} (A).

    \item Model (D): The values $z_{e,\dagger}$=1.7 and $\eta$=10
 were chosen to facilitate a direct comparison between this model and models (A) and (B).\footnote{It is worth noting that increasing 
$\eta$ can render the transition nearly abrupt, resulting in behaviour similar to $\Lambda_{\rm s}$CDM.}   
\end{itemize}

\begin{figure*}[htbp] 
    \centering
    \begin{subfigure}{0.45\textwidth}
        \centering
        \includegraphics[width=\textwidth]{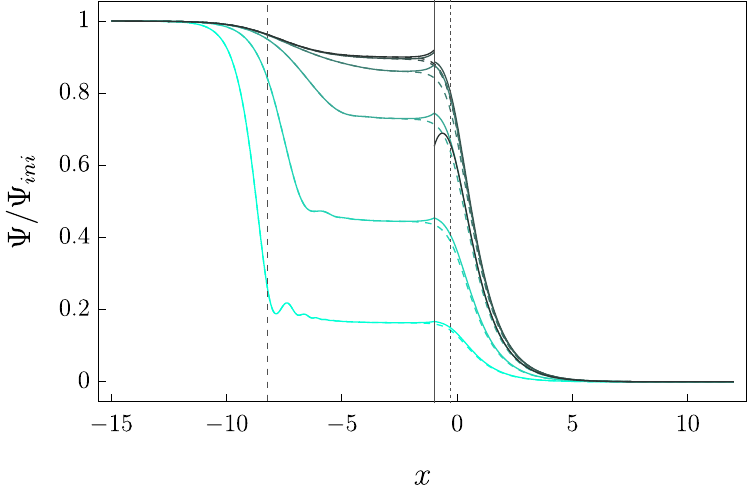}
        \caption*{(A) $\Lambda_{\rm s}$CDM}
    \end{subfigure}
    \hspace{0.05\textwidth}
    \begin{subfigure}{0.45\textwidth}
        \centering
        \includegraphics[width=\textwidth]{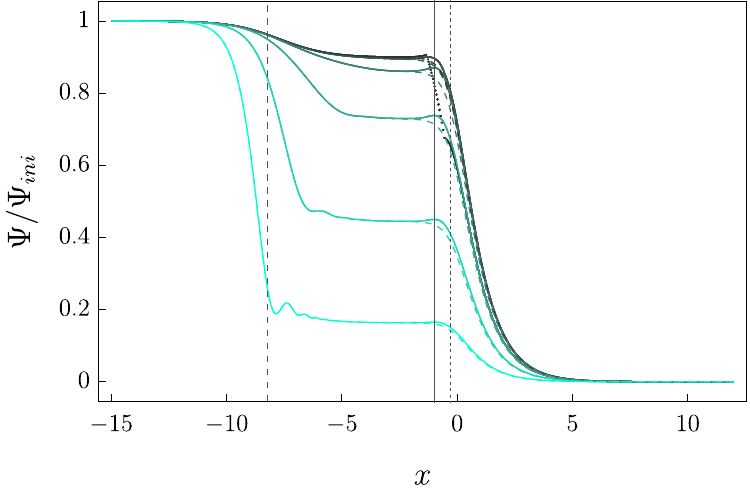}
        \caption*{(B) L$\Lambda$CDM}
    \end{subfigure}

    \vspace{0.5cm} 

    \begin{subfigure}{0.45\textwidth}
        \centering
        \includegraphics[width=\textwidth]{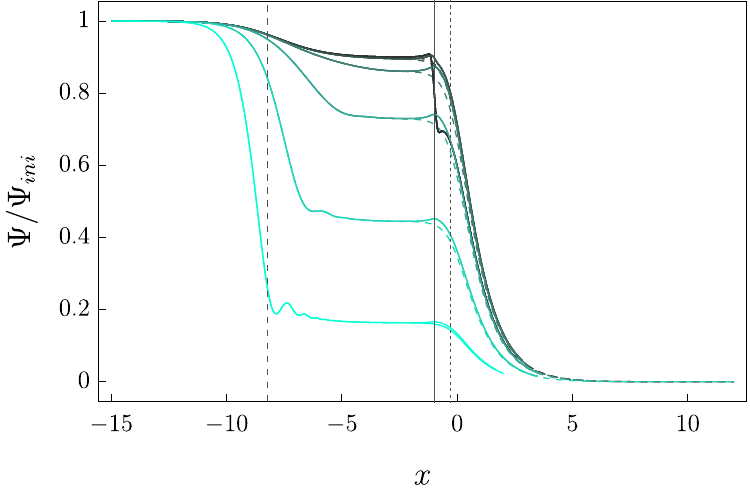}
        \caption*{(C) SSCDM}
    \end{subfigure}
    \hspace{0.05\textwidth}
    \begin{subfigure}{0.45\textwidth}
        \centering
        \includegraphics[width=\textwidth]{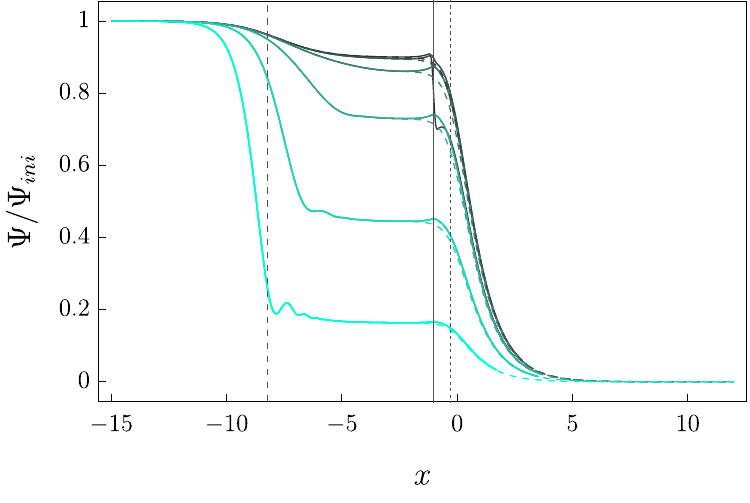}
        \caption*{(D) ECDM}
    \end{subfigure}
    \caption[[Gravitational potential comparison]{\justifying{\textit{
Evolution of the  normalised gravitational potential, $\Psi/\Psi_{\text{ini}}$, for the four different models, shown in comparison with the $\Lambda$CDM scenario. In all panels, solid lines correspond to the model under consideration, while dotted lines represent the $\Lambda$CDM prediction. Each colour denotes a distinct Fourier mode:  
$k = 3.33 \times 10^{-4} \ \text{h} \ \text{Mpc}^{-1}$ (Dark Grey-Green),  
$k = 1.04 \times 10^{-3} \ \text{h} \ \text{Mpc}^{-1}$ (Deep Green),  
$k = 3.27 \times 10^{-3} \ \text{h} \ \text{Mpc}^{-1}$ (Dark Teal),  
$k = 1.02 \times 10^{-2} \ \text{h} \ \text{Mpc}^{-1}$ (Turquoise),  
$k = 3.19 \times 10^{-2} \ \text{h} \ \text{Mpc}^{-1}$ (Teal Green),  
$k = 0.1 \ \text{h} \ \text{Mpc}^{-1}$ (Aqua Blue).  All perturbations are plotted as a function of $x = \ln(a/a_0)$, ranging from well within the radiation-dominated era ($x = -15$) to the distant future ($x = 12$), with $x = 0$ marking the present epoch. The left dashed vertical line indicates radiation-matter equality, the right dashed line marks matter–DE equality, and the solid vertical line corresponds to the redshift at which the DE density changes sign.
}}}
    \label{fig:four-images-psi}
\end{figure*}

In Fig.~\ref{fig:four-images}, we illustrate the evolution of cosmological perturbations from an initial time to a point in the future evolution of the Universe. The four panels display the evolution of the fractional energy density perturbations in matter, $\delta_m$, for the different models under consideration. In each panel, the solid lines represent the results for the respective model, while the dotted lines correspond to those of the $\Lambda$CDM model. For each quantity, we present results for six different wave numbers: $k = 3.33 \times 10^{-4}\, \text{h}\, \text{Mpc}^{-1}$ (dark grey-green), $k = 1.04 \times 10^{-3}\, \text{h}\, \text{Mpc}^{-1}$ (deep green), $k = 3.27 \times 10^{-3}\, \text{h}\, \text{Mpc}^{-1}$ (dark teal), $k = 1.02 \times 10^{-2}\, \text{h}\, \text{Mpc}^{-1}$ (turquoise), $k = 3.19 \times 10^{-2}\, \text{h}\, \text{Mpc}^{-1}$ (teal green), and $k = 0.1\, \text{h}\, \text{Mpc}^{-1}$ (aqua blue). With regard to their evolution, three qualitatively distinct behaviours can be identified, depending on the wave number range:

\begin{itemize}
    \item \textbf{Large $k$} = $0.1 $h Mpc$^{-1}$ (aqua blue) and k = $3.19 \times 10^{-2}$h Mpc$^{-1}$ (teal green).
    \item \textbf{Intermediate $k$} = $1.02 \times 10^{-2}$h Mpc$^{-1}$ (turquoise) and k = $3.27 \times 10^{-3}$h Mpc$^{-1}$ (dark teal).
    \item \textbf{Small $k$} = $1.04 \times 10^{-3}$h Mpc$^{-1}$  (deep green) and k = $3.33 \times 10^{-4}$h Mpc$^{-1}$ (dark gray-green).
\end{itemize}
This classification will be very useful when discussing the evolution of the gravitational potential $\Psi$.

\begin{figure*}[ht]
    \centering
    \begin{subfigure}{0.24\textwidth}
        \centering
        \includegraphics[width=\linewidth]{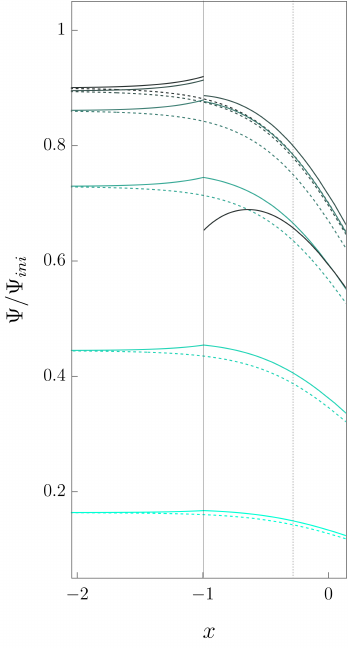}
        \caption*{(A) $\Lambda_{\rm s}$CDM}
    \end{subfigure}
    \begin{subfigure}{0.24\textwidth}
        \centering
        \includegraphics[width=\linewidth]{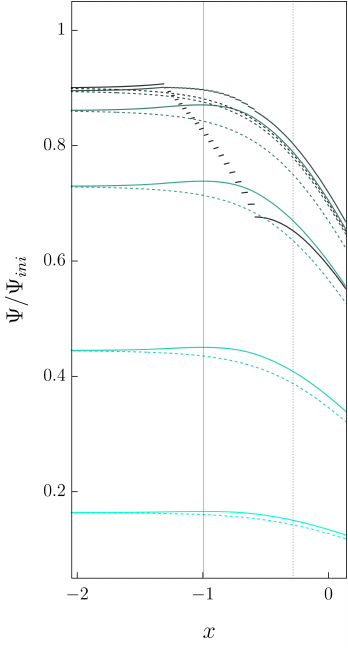}
        \caption*{(B) L$\Lambda$CDM}
    \end{subfigure}
    \begin{subfigure}{0.24\textwidth}
        \centering
        \includegraphics[width=\linewidth]{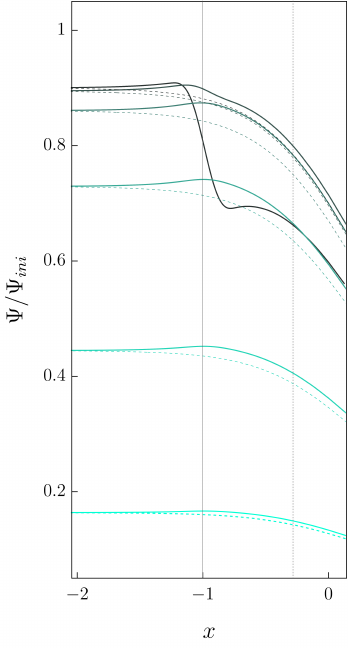}
        \caption*{(C) SSCDM}
    \end{subfigure}
    \begin{subfigure}{0.24\textwidth}
        \centering
        \includegraphics[width=\linewidth]{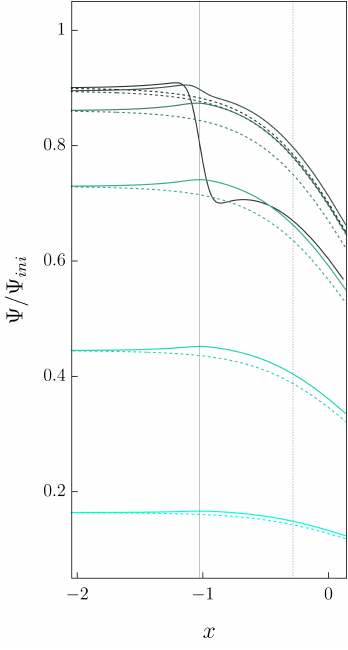}
        \caption*{(D) ECDM}
    \end{subfigure}
    \caption{\justifying{\textit{Gravitational potential $\Psi/\Psi_{ini}$ for the four different models, shown in comparison with the $\Lambda$CDM model. In all panels, solid lines represent the model under study, while dotted lines correspond to the $\Lambda$CDM prediction. The Fourier modes are the same as those used in Fig.~\ref{fig:four-images-psi}. The continuous vertical line indicates the redshift at which the DE density changes sign, and the dotted vertical line denotes the epoch of DE–matter equality.
}}}
    \label{zoompsi}
\end{figure*}

Figure~\ref{fig:four-images} illustrates that the matter perturbations in the sign-switching models closely track those of the $\Lambda$CDM cosmology, with only the smallest $k$ modes exhibiting appreciable deviations; i.e. the modes that enter the horizon around the sign-switching of dark energy. This behaviour is consistently observed across all four models. During the radiation-dominated era, the matter density contrast, $\delta_m$, remains nearly constant for each mode until it crosses the horizon. Upon horizon entry, gravitational collapse triggers the growth of perturbations, which becomes exponential during the matter-dominated epoch. In the subsequent DE-dominated phase, this growth halts, and $\delta_m$ asymptotically tends towards a constant value for each mode. It is also noteworthy that modes with larger wavenumbers, $k$, enter the horizon earlier and consequently reach higher values of $\delta_m$ throughout cosmic evolution.

In Fig.~\ref{fig:four-images-psi}, we present the evolution of the gravitational potential, $\Psi$, normalised to its initial value $\Psi_{\text{ini}}$, for the four models under consideration and the previously discussed wave modes. In each panel, solid lines correspond to the respective model, while dotted lines represent the $\Lambda$CDM prediction. In both the early Universe and the far future, the evolution of $\Psi$ closely follows that of $\Lambda$CDM. However, noticeable deviations arise near the end of the matter-dominated era, when DE begins to exert a significant influence, and extend into the early stages of the DE-dominated epoch. These differences are more clearly discernible in Fig.~\ref{zoompsi}, which offers a magnified view of the gravitational potential’s evolution around the sign-switching epoch.

\begin{itemize}
    \item \textbf{Large $k$}: These modes enter the horizon during the radiation-dominated era, where they undergo decay accompanied by small oscillations, which are further damped shortly before the radiation–matter equality. During the matter-dominated phase, the gravitational potential stabilises at a nearly constant value, typically featuring a modest peak around the sign-switching epoch. This is followed by a second decay phase near the matter–DE equality, lasting for approximately five e-folds. Thereafter, the potential asymptotically approaches a very low constant value, which persists into the far future.

    \item \textbf{Intermediate $k$}: The gravitational potential remains approximately constant throughout the radiation-dominated era. It begins to decay near the epoch of radiation–matter equality and subsequently settles into a second plateau during the matter-dominated phase. The decay occurring during this transition is scale-dependent, predominantly affecting modes with larger wave numbers. Around the sign-switching epoch, a more prominent peak emerges in comparison to the high-$k$ modes, after which the evolution proceeds in a manner analogous to the previously described case.
    
    \item \textbf{Small $k$}: Similar to the intermediate case, the smallest-scale modes remain constant throughout the radiation-dominated era, exhibiting a mild decay around the radiation–matter equality, followed by a second plateau. However, these modes display distinctive features: as shown in Figs.~\ref{zoompsi}\textcolor[rgb]{0.22, 0.67, 0.59}{A} and~\ref{zoompsi}\textcolor[rgb]{0.22, 0.67, 0.59}{B}, the gravitational potential becomes discontinuous at the sign-switching epoch. In Model (A), this manifests as an abrupt drop, while in Model (B), the transition is more gradual due to the greater number of steps in the DE evolution. In contrast, Models (C) and (D), shown in Fig.~\ref{zoompsi}, exhibit a rapid yet continuous decay, reflecting the smoothly varying DE density. After this transition, the evolution of the modes closely resembles that observed in the other scenarios.

\end{itemize}

 \begin{figure*}[htbp] 
    \centering
    \begin{subfigure}{0.45\textwidth}
        \centering
        \includegraphics[width=\textwidth]{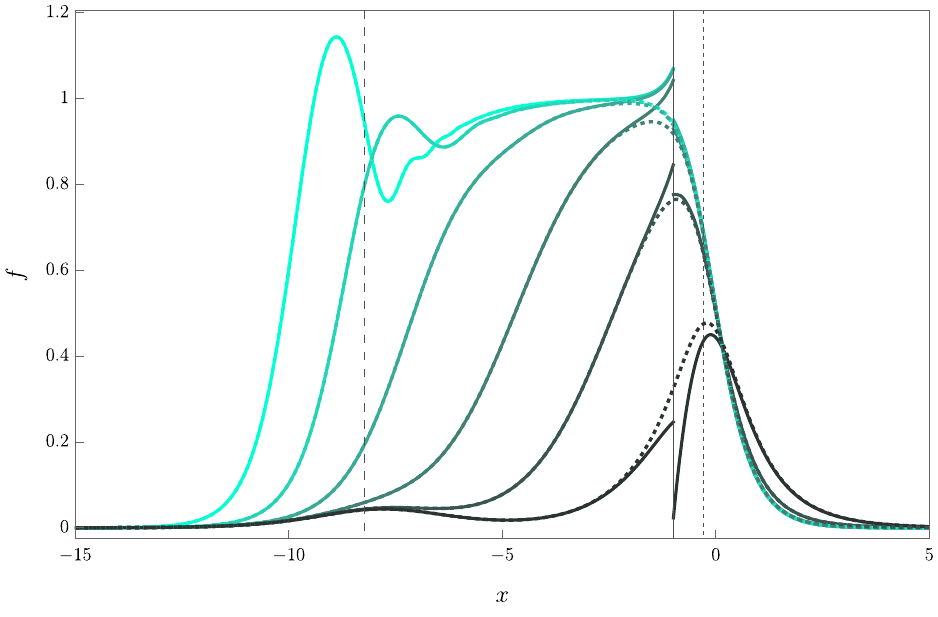}
        \caption*{(A) $\Lambda_{\rm s}$CDM}
    \end{subfigure}
    \hspace{0.05\textwidth}
    \begin{subfigure}{0.45\textwidth}
        \centering
        \includegraphics[width=\textwidth]{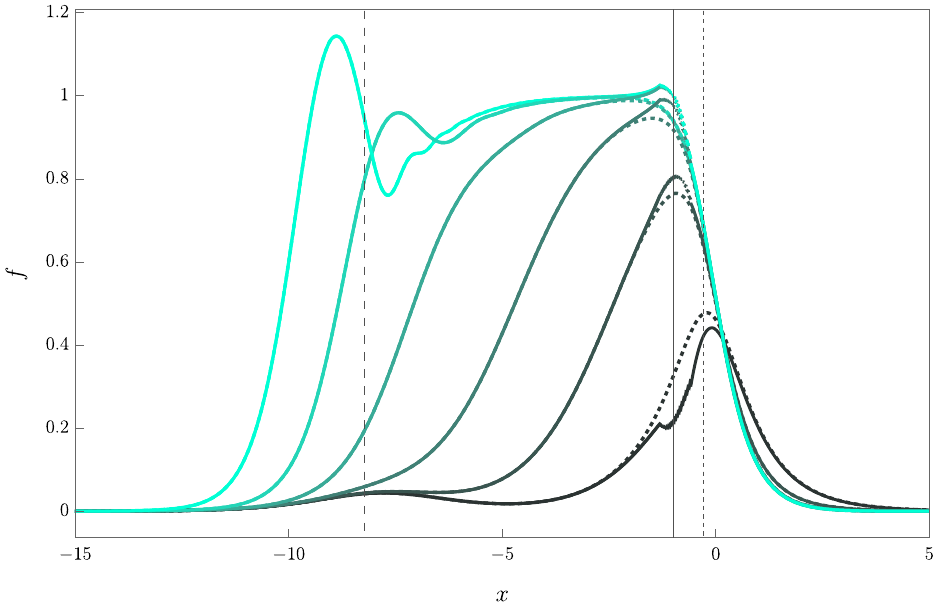}
        \caption*{(B) L$\Lambda$CDM}
    \end{subfigure}

    \vspace{0.5cm} 

    \begin{subfigure}{0.45\textwidth}
        \centering
        \includegraphics[width=\textwidth]{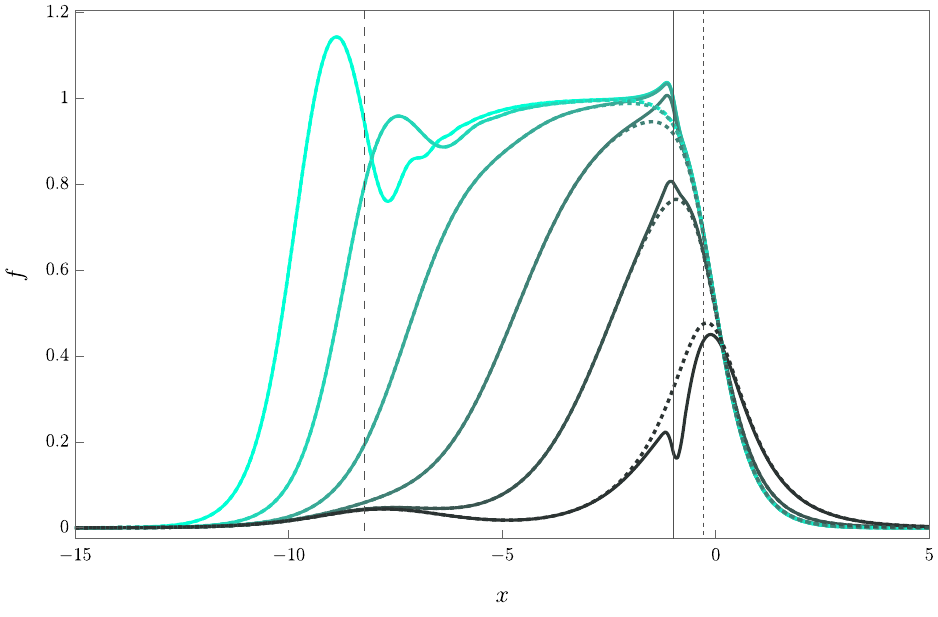}
        \caption*{(C) SSCDM}
    \end{subfigure}
    \hspace{0.05\textwidth}
    \begin{subfigure}{0.45\textwidth}
        \centering
        \includegraphics[width=\textwidth]{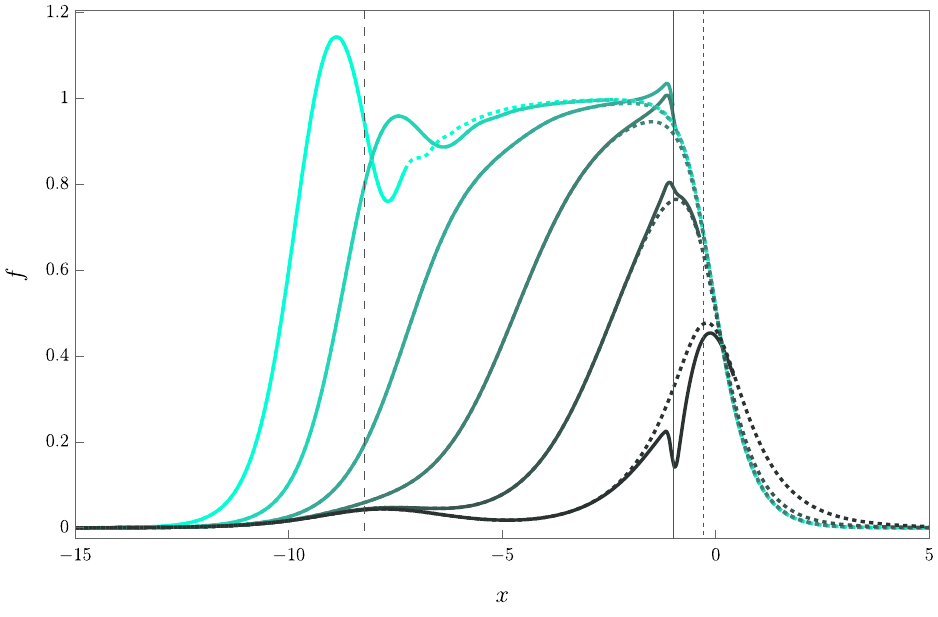}
        \caption*{(D) ECDM}
    \end{subfigure}
    \caption[Growth rate comparison]{\justifying{\textit{
Growth rate $f$ for the four different models, shown in comparison with the $\Lambda$CDM model. Each curve corresponds to a different Fourier mode, following the same convention as in Fig.~\ref{fig:four-images}. The solid lines represent the evolution of the growth rate for each model, while the dashed lines correspond to the $\Lambda$CDM model. The left dashed vertical line indicates the epoch of radiation–matter equality, the right dashed line marks matter–DE equality, and the continuous vertical line denotes the redshift at which the DE density changes sign.
}}}
    \label{fig:four-images-f}
\end{figure*}

\section{COSMOLOGICAL OBSERVABLES}\label{observables}

In this section, we turn our attention to the formation of cosmic structures. Specifically, we examine the growth rate of matter perturbations and investigate how key observables; such as the matter power spectrum and the quantity $f\sigma_8$, can serve as effective tools for constraining our models.

\subsection{Growth rate \label{sec3b}}

One method to constraining the models presented in Section \ref{sec2} involves computing the growth rate of matter, defined as follows \cite{Albarran:2016mdu,Balcerzak:2012ae}:

\begin{equation}
    f=\frac{d(\ln\delta_\textrm{m})}{d(\ln a)}
    \label{eq-f}.
\end{equation}

In Fig.~\ref{fig:four-images-f}, we show the evolution of the growth rate function 
for the four models introduced in Sec.~\ref{sec2}. Initially, all modes remain super-Hubble and nearly constant until they begin to enter the horizon. The modes with the largest $k$ enter first, causing $f$ to increase until it converges to a common value well approximated by $f \approx \Omega_{\mathrm{m}\Lambda}^\gamma$, for modes inside the sub-Hubble regime $(k\gg\mathcal{H})$. For general DE models, this growth index can be modified to $\gamma = 0.55 + 0.05[1 + w_d(z=1)]$ \cite{Linder:2005in}. In this sub-Hubble regime, the dependence on $k$ becomes negligible, enabling a straightforward comparison between the models and $\Lambda$CDM.

At high redshift, all models behave similarly, with modes entering the horizon at comparable times. However, deviations emerge at low redshift. In Fig.~\ref{fig:four-images-f}\textcolor[rgb]{0.22, 0.67, 0.59}{A}, the presence of a negative cosmological constant leads to a continued increase in the growth rate up to the sign-switching redshift. This is due to the attractive nature of the negative DE density, which behaves similarly to an additional matter component, thereby enhancing structure formation. After the sign-switch, the growth rate of the largest modes closely follows the $\Lambda$CDM evolutions . However, the smallest mode exhibits a significantly lower growth rate, seeming like a suppression of structure formation on those scales. This is because these modes correspond to very large scales. At such scales, the modes have only recently entered the horizon and have not had sufficient time to grow significantly. At late times, once DE dominates, all models converge again in their evolution. In Fig.~\ref{fig:four-images-f}\textcolor[rgb]{0.22, 0.67, 0.59}{B}, a similar behaviour is observed, though the transition is smoother due to the gradual variation in DE density during the sign-switch. 
This smooth transition moderates the enhancement of structure growth compared to the abrupt $\Lambda_{\rm s}$CDM case, although it still predicts more structure than $\Lambda$CDM. Finally, models (C) and (D); i.e. those with continuous DE densities, exhibit a small bump in the growth rate just prior to the sign-switch. At this point, the DE density has become sufficiently negative to noticeably influence the dynamics. This negative energy density, sometimes referred to as \textit{phantom matter}  in the literature \cite{Gomez-Valent:2024tdb,Gomez-Valent:2024ejh}, acts as an additional attractive component, further boosting structure formation until it changes sign. The \textit{phantom matter} term refers to a peculiar form of DE which, in marked contrast to conventional phantom DE, is distinguished by having a positive pressure ($p_\text{d} > 0$) at the cost of a negative energy density ($\rho_\text{d} < 0$). Following the transition, the now-positive DE density, combined with the declining matter density, leads to a reduction in the growth rate, eventually aligning with the $\Lambda$CDM behaviour.

In \cite{Nguyen:2023fip}, a suppressed growth rate of large-scale structure during the dark energy-dominated era was identified using Planck 2018 data \cite{Planck:2018vyg} and redshift-space distortion (RSD) measurements \cite{Avila:2022xad}. The findings indicated a larger growth index, $\gamma = 0.639$, compared to the commonly accepted value of $\gamma = 0.55$, highlighting a tension in the growth index $\gamma$ within the $\Lambda$CDM model. Using a similar methodology, the authors of \cite{Escamilla:2025imi} constrained the growth index for the $\Lambda_{\rm s}$CDM model with the same datasets, demonstrating that this model can also alleviate the $\gamma$ tension. Once the corresponding observational fits are performed in \cite{fit}, we likewise expect our models to reduce this tension.

\subsection{Matter power spectrum \label{sec3c}}
The distribution of matter across different scales in the Universe can be analysed through the matter power spectrum. In the Newtonian gauge, it is defined as \cite{Albarran:2016mdu}
\begin{equation}
    P = |\delta_m - 3\mathcal{H}v_m|^2.
\end{equation}

\begin{figure}[htbp]
    \centering
    \includegraphics[width=1\linewidth]{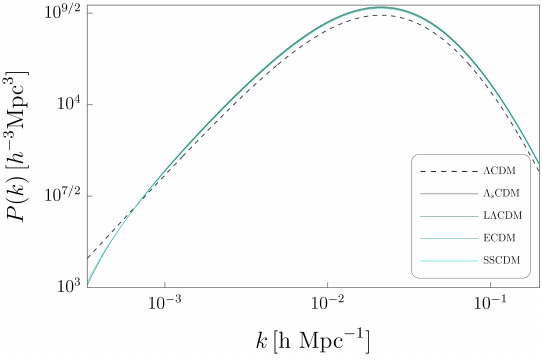}
    \caption{\justifying{\textit{Matter power spectrum as a function of the Fourier modes. The dashed black line corresponds to the $\Lambda$CDM model, while the green curves represent the sign-switching models. It can be observed that the sign-switching models largely overlap with one another.}}}
    \label{fig:power}
\end{figure}

In Fig.~\ref{fig:power}, we depict the matter power spectrum for wave-numbers in the range $k \in [0.000334, 0.2]~\text{h}\,\mathrm{Mpc}^{-1}$. As observed previously in the case of the growth rate, the larger $k$-modes (corresponding to smaller scales) tend to enhance structure formation, whereas the smaller $k$-modes appear comparatively suppressed. However, this trend  could change upon performing a full cosmological fit. Indeed, Fig.~\ref{fig:four-images-f} indicates an apparent overall enhancement in the growth of structure compared to the $\Lambda$CDM scenario. Of course, this is a preliminary result and requires a further step, namely fitting the models to the observational data \cite{fit}.

\subsection{\texorpdfstring{$f\sigma8$}{fs8} distribution \label{sec3d}}

\begin{figure*}[htbp] 
    \centering
    \begin{subfigure}{0.45\textwidth}
        \centering
        \includegraphics[width=\textwidth]{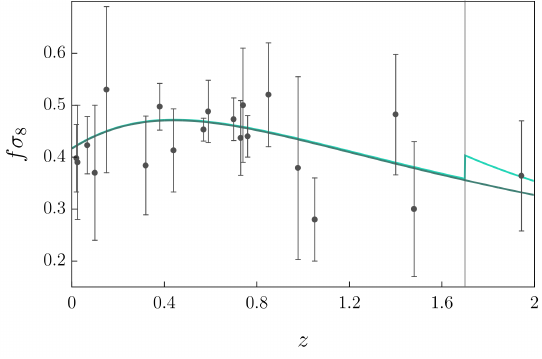}
        \caption*{(A) $\Lambda_{\rm s}$CDM}
    \end{subfigure}
    \hspace{0.05\textwidth}
    \begin{subfigure}{0.45\textwidth}
        \centering
        \includegraphics[width=\textwidth]{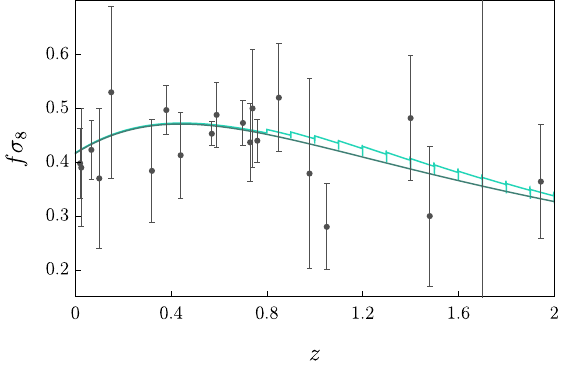}
        \caption*{(B) L$\Lambda$CDM}
    \end{subfigure}

    \vspace{0.5cm} 

    \begin{subfigure}{0.45\textwidth}
        \centering
        \includegraphics[width=\textwidth]{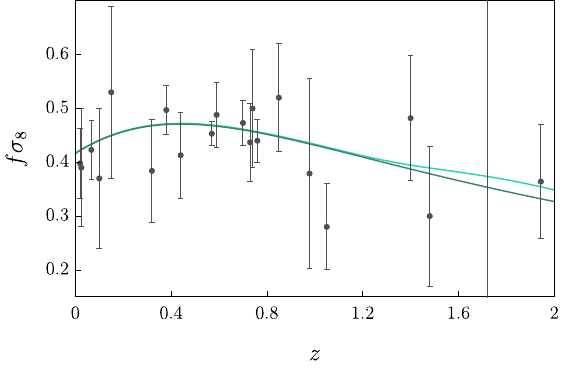}
        \caption*{(C) SSCDM}
    \end{subfigure}
    \hspace{0.05\textwidth}
    \begin{subfigure}{0.45\textwidth}
        \centering
        \includegraphics[width=\textwidth]{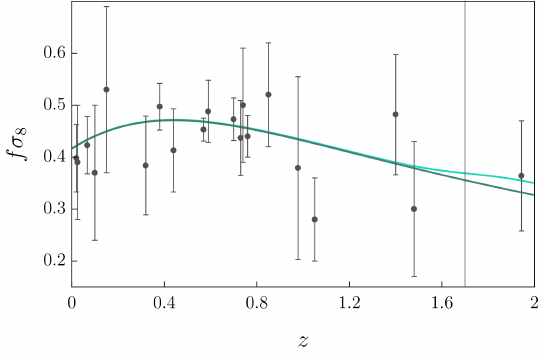}
        \caption*{(D) ECDM}
    \end{subfigure}

    \caption{\justifying{\textit{Evolution of $f \sigma8$ for redshift between $z=0$ and $z=2$ against the available data points of Ref. \cite{Avila:2022xad} . The light green lines are the respective models of study while the dark green one is $\Lambda$CDM. 
    }}}
    \label{fig:four-images-fs8b}
\end{figure*}
Another key observable frequently appearing in cosmological data is the product $f\sigma_8$, rather than the growth rate $f$ alone. Here, $\sigma_8$ denotes the root mean square of the matter density fluctuations within spheres of radius $8\,\text{h}^{-1}$Mpc, and serves as a normalisation parameter for the matter power spectrum. The main advantage of considering the combination $f\sigma_8$ lies in its ability to mitigate the degeneracy between $\sigma_8$ and the linear galaxy bias $b$, which relates the dark matter perturbations to the observed galaxy density fluctuations. The time evolution of $\sigma_8$ can be computed via the following expression:

\begin{equation}
\sigma_8(z,k_{\sigma8})=\sigma_8(0,k_{\sigma8})\frac{\delta_\textrm{m}(z,k_{\sigma8})}{\delta_\textrm{m}(0,k_{\sigma8})}.
    \label{eq-sigma8}
\end{equation}
Here, $k_{\sigma_8} = 0.125$\,h\,Mpc$^{-1}$ denotes the wavenumber corresponding to a physical scale of $8\,\text{h}^{-1}$\,Mpc. We compute the evolution of $f\sigma_8$ by solving Eqs.~(\ref{dengen}) and (\ref{velgen}), in conjunction with relations (\ref{eq-f}) and (\ref{eq-sigma8}). For all models under consideration, we adopt the Planck 2018 value $\sigma_8(0, k_{\sigma_8}) = 0.8120$ \cite{Planck:2018vyg} as the present-day normalisation. The resulting predictions will be compared both with those of the $\Lambda$CDM model and with observational data compiled in Ref.~\cite{Avila:2022xad}.

Combining all these elements, we construct Fig.~\ref{fig:four-images-fs8b}, which presents the evolution of $f\sigma_8$ relative to $\Lambda$CDM for the four models under study, over the redshift interval $z \in (0,2)$. In each panel, the predictions of our models are shown in light green, while those of $\Lambda$CDM appear in a darker green. We find that all models remain within the observational uncertainties reported in Ref.~\cite{Avila:2022xad} across most redshifts. Notably, each of the four models exhibits an improved match to the highest redshift value derived from quasars, EBOSS data \cite{Planck:2018vyg}, lying closer to it than the $\Lambda$CDM model. In Fig.~\ref{fig:four-images-fs8b}\textcolor[rgb]{0.22, 0.67, 0.59}{B}, we observe a generalised version of the behaviour shown in panel Fig.~\ref{fig:four-images-fs8b}\textcolor[rgb]{0.22, 0.67, 0.59}{A}, featuring multiple discrete transitions rather than a single abrupt step. In Figs.~\ref{fig:four-images-fs8b}\textcolor[rgb]{0.22, 0.67, 0.59}{C} and ~\ref{fig:four-images-fs8b}\textcolor[rgb]{0.22, 0.67, 0.59}{D} we can observe the transition from higher $f\sigma8$ values to the actual $\Lambda$CDM values as the DE density transitions from negative to positive values. In all four models, an enhancement of structure formation is observed in the vicinity of the sign-switching event.

\section{CONCLUSION \label{sec5}}

Motivated by the compelling $\Lambda_{\rm s}$CDM model, also referred to as model (A), we have carried out a comprehensive perturbative analysis of four dynamical DE scenarios characterised by sign-switching behaviour in the cosmological constant. In particular, we have reviewed a generalised ladder-step DE density profile, labelled model (B) or L$\Lambda$CDM, and studied two additional frameworks, models (C) and (D) or SSCDM and ECDM, respectively, in which the transition from negative to positive DE density occurs smoothly.

After outlining the relevant aspects of cosmological perturbation theory and specifying suitable initial conditions, we numerically solved Eqs.~(\ref{perturb}) and (\ref{fieldk}) to trace the evolution of linear matter perturbations across all four models. Our findings indicate that, at the level of matter perturbations, the models remain largely indistinguishable from $\Lambda$CDM, with only minimal deviations.

We then examined the evolution of the gravitational potential. While the early and late-time behaviours remain close to those predicted by $\Lambda$CDM, noticeable differences arise around the epoch of matter–DE equality. In particular, models (A) and (B), which feature discontinuous transitions in the total EoS , exhibit abrupt features in the gravitational potential, especially for the smallest modes ($k < 10^{-3}$h Mpc$^{-1}$). This behaviour is expected, since the attractive nature of the negative-density DE, whose relative density becomes significant at these redshifts, acts effectively as \textit{phantom matter}, increasing the contribution of components with matter-like characteristics. Following the transition, DE resumes its repulsive character with negative pressure, resulting in a suppression of the gravitational potential. By contrast, the smooth-transition models (C) and (D) avoid such discontinuities but display a rapid decay in the gravitational potential for small-scale modes, after which it stabilises close to zero, as in the standard model. Nevertheless, for the largest modes, an enhancement relative to $\Lambda$CDM is observed around the sign-switching epoch.

Our investigation proceeded with the analysis of cosmological observables, including the growth rate $f$, the matter power spectrum, and the $f\sigma_8$ parameter. We observed a mild enhancement in structure formation relative to $\Lambda$CDM across all four models, most notably in the continuous scenarios (C) and (D), where the negative energy density component—often interpreted as  \textit{phantom matter}—acts to amplify growth. The evolution of $f\sigma_8$ in particular shows improved agreement with low-redshift RSD data, especially when compared to the biggest derived data point.

In summary, both the background-level analysis \cite{Bouhmadi-Lopez:2025ggl} and the perturbative results presented here support the theoretical consistency and observational plausibility of sign-switching DE models. We have observed that such models favour an enhancement in structure growth and formation as favoured by observational data. Nevertheless, a robust statistical comparison with cosmological datasets is essential to further constrain the parameter space. Such an analysis, including a full likelihood exploration and model fitting, is deferred to forthcoming work \cite{fit}.

\section*{Acknowledgements}

The authors are grateful to \"{O}zg\"{u}r Akarsu, Carlos G. Boiza, Hsu-Wen Chiang, Nihan Kat{\i}rc{\i}  and Thomas Broadhurst for discussions and insights on the current project. M. B.-L. is supported by the Basque Foundation of Science Ikerbasque. 
Our work is supported by the Spanish Grant PID2023-149016NB-I00 funded by (MCIN/AEI/10.13039/501100011033 and by “ERDF A way of making Europe). This work is also supported by the Basque government Grant No. IT1628-22 (Spain) and in particular B. I. U. is funded through that Grant.

\bibliographystyle{elsarticle-num} 
\bibliography{bibliografia(marked)}

\end{document}